\documentclass[twocolumn,prl,footinbib]{revtex4}
\usepackage{amsfonts, amssymb}
\usepackage{graphicx, epsfig,bm}
\usepackage{color}

\newcommand{\be}{\begin{equation}}
\newcommand{\ee}{\end{equation}}
\newcommand{\bea}{\begin{eqnarray}}
\newcommand{\eea}{\end{eqnarray}}

\def\fun#1#2{\lower3.6pt\vbox{\baselineskip0pt\lineskip.9pt
        \ialign{$\mathsurround=0pt#1\hfill##\hfil$\crcr#2\crcr\sim\crcr}}}

\newcommand\lsim{\mathrel{\rlap{\lower4pt\hbox{\hskip1pt$\sim$}}
    \raise1pt\hbox{$<$}}}
\newcommand\gsim{\mathrel{\rlap{\lower4pt\hbox{\hskip1pt$\sim$}}
    \raise1pt\hbox{$>$}}}

\def\dslash{\not{\hbox{\kern-2pt $\partial$}}}
\def\Dslash{\not{\hbox{\kern-4pt $D$}}}
\def\Oslash{\not{\hbox{\kern-4pt $O$}}}
\def\Qslash{\not{\hbox{\kern-4pt $Q$}}}
\def\pslash{\not{\hbox{\kern-2.3pt $p$}}}
\def\kslash{\not{\hbox{\kern-2.3pt $k$}}}
\def\qslash{\not{\hbox{\kern-2.3pt $q$}}}

 \newtoks\slashfraction
 \slashfraction={.13}
 \def\slash#1{\setbox0\hbox{$ #1 $}
 \setbox0\hbox to \the\slashfraction\wd0{\hss \box0}/\box0 }

\def\ee{\end{equation}}
\def\be{\begin{equation}}

\begin{document}
\setlength{\unitlength}{1mm}

\title{Beyond Two Dark Energy Parameters} \author{
Devdeep Sarkar$^1$, Scott Sullivan$^1$, Shahab Joudaki$^1$, Alexandre Amblard$^1$, Daniel
E. Holz$^{2,3}$,  and Asantha Cooray$^1$} \affiliation{$^1$Department
of Physics and Astronomy,  University of California, Irvine, CA 92697}
\affiliation{$^2$Theoretical    Division,   Los   Alamos   National
Laboratory,  Los  Alamos, NM  87545}  \affiliation{$^3$Department  of
Astronomy \& Astrophysics, University of Chicago, Chicago, IL 60637}

\date{\today}%

\begin{abstract}

Our  ignorance  of  the  dark  energy  is  generally  described  by  a
two-parameter  equation of  state.  In  these approaches  a particular
{\it  ad hoc}  functional form  is assumed,  and only  two independent
parameters   are  incorporated.    We  propose   a  model-independent,
multi-parameter  approach to fitting  the dark  energy, and  show that
next-generation surveys will constrain  the equation of state in three
or  more independent  redshift  bins to  better  than $10$\%.   Future
knowledge  of  the  dark   energy  will  surpass  two  numbers  (e.g.,
[$w_0$,$w_1$]  or  [$w_0$,$w_a$]),  and  we propose  a  more  flexible
approach to the analysis of present and future data.
\end{abstract}
\bigskip

\maketitle

Standard candles such as Type Ia supernovae (SNe), as well as standard
rulers such  as the cosmic  microwave background (CMB) and  the baryon
acoustic oscillation  (BAO) scale, are currently  the preferred probes
of the expansion  history of the Universe \cite{Rieetal04,Tonryetal03,
Spergel06,  Eis05}. By determining  distances at  cosmological scales,
these  probes  have  firmly  established  that the  expansion  of  the
universe is accelerating~\cite{perlmutter-1999,Rie06,Wood,Astier}.  It
is now believed that a mysterious dark energy component with an energy
density  $\sim$70\% of  the total  energy density  of the  universe is
responsible for this accelerated expansion.  The underlying physics of
dark energy  remains obscure \cite{Padmanabhan}  and understanding the
acceleration has become one  of the foremost challenges in fundamental
physics.

In an attempt to discriminate observationally between differing models
of dark  energy, it is useful  to parameterize the dark  energy by its
equation  of  state (EOS),  encapsulating  the  ratio  of pressure  to
density.  When  model fitting data,  it is generally assumed  that the
dark energy  EOS follows a certain  predetermined evolutionary history
with  redshift,  $w(z)$.  Common  parameterizations  include a  linear
variation,  $w(z)=w_0+w_z  z$ \cite{CooHut99},  or  an evolution  that
asymptotes to a constant $w$ at high redshift, $w(a)=w_0 + w_a (1-a)$,
with $a$ the scale factor \cite{ChevPol, Linder}.

Fitting data to an assumed functional form leads to possible biases in
the determination of properties of  the dark energy and its evolution,
especially  if  the true  behavior  of  the  dark energy  EOS  differs
significantly from the  assumed form~\cite{Gerke}.  The issues related
to  model-dependent studies  of  the  dark energy  EOS  are a  greater
problem for the high  precision datasets expected from next-generation
cosmological experiments, including distance measurements from a Joint
Dark Energy Mission (JDEM
\footnote{http://universe.nasa.gov/program/probes/jdem.html}).

Instead of  using a parameterized form  for $w(z)$, one  can utilize a
variant of  principal component analysis  \cite{HutSta03} to establish
the EOS  without relying  on a specific  parameter description  of the
underlying  redshift  evolution.   This  was  applied  by  Huterer  \&
Cooray~\cite{HutCoo} to a set  of early supernova data.  More recently
Riess et al. \cite{Rie06} used the  same approach to analyze a new set
of $z  > 1$  SNe from  the Hubble Space  Telescope, while  an analysis
involving a larger combined  dataset \cite{Davis07} has been presented
in   Sullivan   et   al.   \cite{sullivan07}.   Here   we   use   this
model-independent approach  to study the  extent to which  future data
will constrain  dark energy.  We find that  more than  two independent
parameters of the EOS  can be determined with next-generation surveys.
Our results argue against claims in the literature that
next-generation surveys can only determine two parameters of the EOS
as a function of redshift \cite{LinHut} and we motivate a
model-free approach to study dark energy.

To encapsulate  the range of  possible future dark energy  surveys, we
consider six different data scenarios  and generate mock data for each
one      assuming      a      flat      $\Lambda$CDM      cosmological
model~\cite{Spergel06}. Our datasets are: 

\begin{itemize}
\item Case A:  A catalog of 200 SNe  uniformly distributed in redshift
  out to $z=1.8$;  in addition, two BAO distance  estimates at $z=0.2$
  and $z=0.35$, with 6\%  and 4.7\% uncertainties, respectively.  This
  case   approximates    the   current   state   of    the   data   in
  SNe~\cite{Rie06,Astier,Davis07} and BAOs~\cite{Eis05,Percival}.
\item Case  B: A catalog  of 300 SNe  uniformly distributed out  to $z
  =0.1$, as  expected from ground-based  low redshift samples,  and an
  additional 2,000 SNe uniformly distributed in the range $0.1<z<1.8$,
  as expected from {\it  JDEM} or similar future surveys~\footnote{Our
  results  are  insensitive  to  the  precise shape  of  the  redshift
  distributions.   Furthermore, a mission  like {\it  SNAP} (SuperNova
  Acceleration Probe) would find greater  than 2,000 SNe, and a subset
  of  the SNe  with a  uniform redshift  distribution is  expected for
  analysis.}.  In addition to the  two BAO distances described in Case
  A,  five additional BAO  constraints at  $z =  [0.6, 0.8,  1.0, 1.2,
  3.0]$ with  corresponding fiducial survey precisions  of $[4.3, 3.2,
  2.3, 2.0, 1.2]$\% (V1N1 from~\cite{Seo03}).
\item Case  C: The same SN dataset  as described in case  B, the seven
  BAO estimates as described in case  B, and, in addition, ten new BAO
  constraints expected  from a  proposed JDEM mission  by NASA  or ESA
  concentrating  primarily on  BAO measurements,  such as  {\it ADEPT}
  (Advanced Dark Energy Physics  Telescope).  These BAO estimates have
  precision         (in         $D_V$         \cite{Eis05})         of
  $[0.36,0.33,0.34,0.33,0.31,0.33,0.32,0.35,0.37,\\0.37]$\%        from
  $z=1.05$ to 1.95 in steps of 0.05 \cite{Eisen}.
\item Case  D: A  dataset of 10,000  SNe uniformly distributed  out to
  $z=2$. In addition, seven BAO constraints as in case B, but assuming
  stronger   accuracies   (V5N5    of~\cite{Seo03})   for   the   five
  higher-redshift BAO constraints: $[1.9, 1.5, 1.0, 0.9, 0.6]$\% at $z
  = [0.6, 0.8, 1.0, 1.2, 3.0]$.
\item  Case  E:  The  10,000  SN  dataset along  with  the  seven  BAO
  constraints  as  described in  case  D  and  the additional  10  BAO
  constraints expected  from space-based missions as  decribed in case
  C.

\item Case F: The SN dataset as described in case A
 combined with the  BAO estimates in case  E.
\end{itemize}

For each of  the cases listed above we create mock  catalogs of SN and
BAO  observations.   For  each  individual  SN we  simulate  a  random
distance  modulus  $m(z)-{\mathcal M}$  consistent  with our  fiducial
cosmological model.   In cases A  to C, we  bin the Hubble  diagram at
$z>0.1$ into 50 redshift bins, while  the 10,000 SNe sample in cases A
to F is binned into 500  bins.  The error in distance modulus for each
SN   bin  is  given   by  $\sigma_m=\sqrt{(\sigma_{\mbox{\footnotesize
int}}/\sqrt{N_{\mbox{\footnotesize   bin}}})^2+\delta   m^2}$,   where
$\sigma_{\mbox{\footnotesize  int}}=0.1\,\mbox{mag}$ is  the intrinsic
error  for each SN~\cite{kim},  $N_{\mbox{\footnotesize bin}}$  is the
number of SNe  in the redshift bin, and $\delta  m$ is the irreducible
systematic  error.  We  take the  systematic  error to  have the  form
$\delta m=0.02(0.1/\Delta  z)^{1/2} (1.7/z_{\mbox{\footnotesize max}})
(1 + z)/2.7$, where  $z_{\mbox{\footnotesize max}}$ is the redshift of
the furthest SNe, and $\Delta z$ is the width of the relevant redshift
bin.  This is equivalent to  the form in \cite{LinHut03}. In addition,
we     include     the     effects    of     gravitational     lensing
magnification~\cite{Frieman,HolzWald98,Wang00},   although  the  noise
from lensing is expected to be  small for large datasets due to sample
averaging  \cite{HolzLinder,Sarkar}.  We make  use of  the probability
distribution function  for lensing  magnification from Wang,  Holz, \&
Munshi~\cite{WangHolzMunshi} at $z \geq 0.6$, while for SNe at $z<0.6$
we   approximate  the   lensing   by  a   Gaussian  distribution   for
magnification with dispersion $0.093z$ magnitudes, as given in Holz \&
Linder~\cite{HolzLinder}.

We analyze a set of 10  independent mock catalogs for the data cases B
and C, to  account for random variations in the  estimates of the EOS.
When model  fitting each of  the mock samples,  we follow~\cite{Rie06}
and  marginalize  over  a  prior  in $\Omega_mh$  ($0.213  \pm  0.023$
\cite{Tegmark04}), a prior in $H_0$ (72 $\pm$ 8 km s$^{-1}$ Mpc$^{-1}$
\cite{key01}),  and a  prior in  the distance  to the  last scattering
surface  at $z=1089$  ($R=1.71 \pm  0.03$ \cite{Wang}).   In  order to
account for uncertainty in the absolute magnitude, we also marginalize
over the  nuisance parameter,  $\mathcal M$, with  a uniform  prior of
-0.6 to  0.6 \cite{Wood}.  To  explore the importance of  our fiducial
assumption of  a flat universe,  for cases C  and E we also  explore a
curvature  prior,  with  a  1$\sigma$  uncertainty  on  $\Omega_k$  of
$0.0032$ \cite{Smith}.

We refer the reader to  Sullivan et al.  \cite{sullivan07} for details
of our  approach.  We make use  of a modified version  of the publicly
available              {\it              wzbinned}\/              code
\footnote{http://www.cooray.org/sn.html}, which analyzes observational
cosmological  data via a  Markov Chain  Monte Carlo  (MCMC) likelihood
approach   to   estimate   $w(z)$   in   redshift   bins.    In   this
parameterization, the  EOS is  taken to be  constant in  each redshift
bin, but can vary from bin to bin. We take a total of six bins between
$z=0$ and $z=3$ and fix the EOS at higher redshift to a constant value
of $-1$  out to the  CMB at $z=1089$.   In addition, we also  impose a
prior  on our  furthest (6th)  bin: $-5  \leq w_6  \leq 0$.   This bin
remains largely unconstrained  by the data cases we  have studied, and
the prior  facilitates convergence. In what follows  we will generally
omit this bin, as it  is poorly constrained. The redshift intervals of
the first five bins are listed in Table~1 and the intervals are chosen
so that the error on $w(z_i)$ is spread evenly across all bins.

The integration  to higher redshifts causes  correlations between lower-
and higher-redshift $w_i(z)$ bins, and this must be taken into account
in the subsequent analysis. These correlations are encapsulated in the
covariance matrix, which can be generated by taking the average of the
Markov  chain, $C  = \left\langle  {\bf  w} {\bf  w}^T \right\rangle  -
\left\langle    {\bf    w}\right\rangle    \left\langle   {\bf    w}^T
\right\rangle$, where  ${\bf w}$ captures  estimates of $w_i(z)$  as a
vector.   This covariance  matrix is  non-diagonal,  with correlations
between  adjacent  bins  that  slowly  decrease  with  increasing  bin
separation.

Instead of restricting ourselves  to correlated values of $w_i(z)$, we
follow  Huterer  \&   Cooray~\cite{HutCoo}  and  decorrelate  the  EOS
estimates.   This  is  achieved  by  changing  the  basis  through  an
orthogonal  matrix rotation that  diagonalizes the  inverse covariance
matrix.  The Fisher  matrix  ${\bf  F} \equiv  C^{-1}$  is then  ${\bf
F}={\bf  O^T} \Lambda  {\bf  O}$  where the  matrix  $\Lambda$ is  the
diagonalized   inverse  covariance  of   the  transformed   bins.  The
uncorrelated parameters are then  defined by the rotation performed by
the orthogonal matrix: ${\bf q} = {\bf O} {\bf w}$.

There is a freedom of choice  in the orthogonal matrix used to perform
this   transformation.   We   follow   the   approach   advocated   in
~\cite{HutCoo}   and  write  the   weight  transformation   matrix  as
$\tilde{{\bf W}} = {\bf O}^T \Lambda^{\frac{1}{2}} {\bf O}$, where the
rows are summed such that the  weights from each band add up to unity.
This choice  ensures we have mostly positive  contributions across all
bands, an  intuitively pleasing  result.  We apply  the transformation
$\tilde{{\bf W}}$ to  each link in the Markov chain  to generate a set
of independent, uncorrelated  measures of the probability distribution
of the  EOS in each bin  as determined by the  observables.  We denote
these uncorrelated bins as  $\tilde{{\bf w}}=\tilde{{\bf W}} {\bf w}$.
When  discussing  our  results,  we  will  generally  refer  to  these
uncorrelated   estimates.

\begin{figure}[!t]
\epsfxsize=3.4in
\centerline{\epsfbox{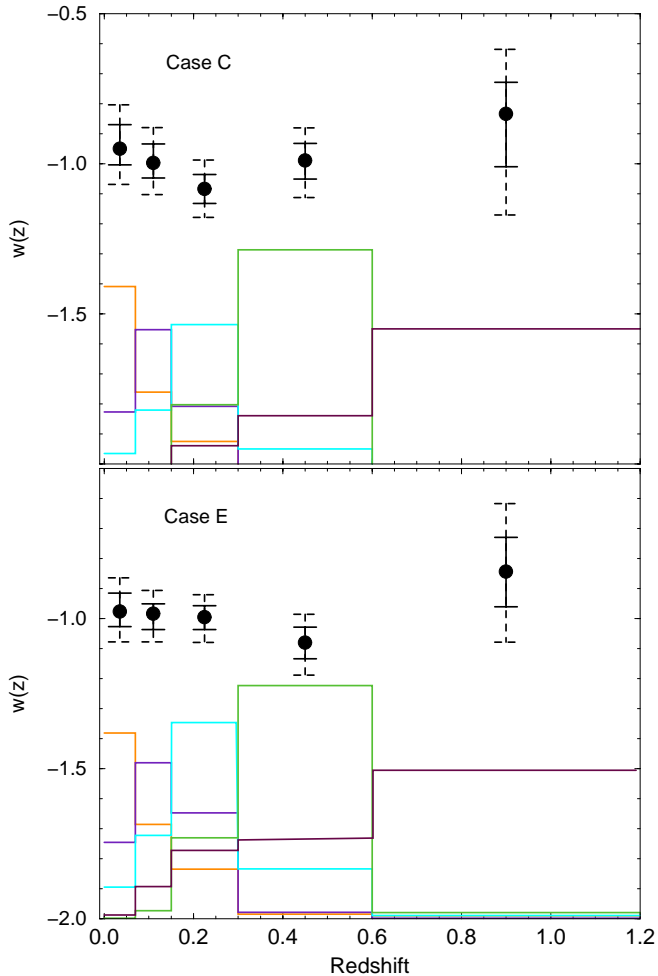}}
\caption{Uncorrelated binned estimates of  the EOS from a typical mock
sample generated  for Case  C (top panel)  and Case E  (bottom panel).
The error  bars show 1$\sigma$ and 2$\sigma$  uncertainties with solid
and dashed lines, respectively.  At the bottom, in separate colors, we
show  the window function  for each  of the  independent, decorrelated
bins.  The  overlap  in   window  functions  represents  the  relative
contribution  from  adjacent   redshifts  bins  to  each  uncorrelated
$w(z_i)$ estimate.  }
\label{figure1}
\end{figure}

The results of our analysis are summarized in Table~\ref{table1}.  For
each of  the observational scenarios  we give the $1\sigma$  errors on
the determination  of $w$ in  5 uncorrelated redshift  ``bins''. Since
the redshift  bins are decorrelated,  they now leak into  one another;
the new window  functions are shown in Fig.~1  where we illustrate the
expected errors in  $w(z_i)$ for two mock samples from  Cases C and E.
For Cases B  and C we also list the relative  dispersion of the binned
$w(z)$ errors  determined by  analyzing 10 independent  datasets.  For
Case C we find that the  first four bins have average errors of 0.067,
0.056, 0.048, and 0.059.  Analyzing several independent mock catalogs,
we   found   the  standard   deviation   to  be   [3.3,1.6,2.5,2.9]\%,
respectively for the first four  bins, relative to the mean error.  In
the last  column of Table~\ref{table1}, we specify  the number ($N_P$)
of  {\em  independent}  EOS   parameters  that  could  be  determined,
according to  our analysis, to an  accuracy better than  10\%. This is
the  same  criteria  used  previously to  argue  that  next-generation
surveys can determine at most two parameters of the EOS \cite{LinHut}.

\begin{table}[!t]\footnotesize
\begin{center}
\begin{tabular}{r|ccccc|c}
\hline
Bin      & $w_1(z)$ & $w_2(z)$   & $w_3(z)$  & $w_4(z)$ & $w_5(z)$ &  $N_P$ \\
z range  & 0-0.07   & 0.07-0.15  & 0.15-0.3  & 0.3-0.6  & 0.6-1.2   &     \\
\hline \hline
Case A     &  0.192       & 0.151   & 0.114     &  0.130     & 0.192    &  0 \\
\hline
Case B      & 0.077       & 0.066       & 0.061       & 0.084    & 0.230  & 4 \\
           & (1.9\%)     & (2.2\%)  & (1.8\%)  & (5.2\%)  & (12.8\%) &    \\
\hline
Case C     &  0.067       & 0.056   & 0.048     &  0.059     & 0.153  &   4 \\
           & (3.3\%)     & (1.6\%)  & (2.5\%)  & (2.9\%)  & (8.1\%) &  \\
with $\Omega_K$ & 0.073  & 0.060  & 0.053  & 0.065  & 0.179 &  4  \\
\hline
Case D      &  0.059       & 0.051   & 0.048     &  0.072    & 0.193   & 4 \\
\hline
Case E      & 0.055      &  0.044      & 0.040       &  0.052   &  0.116 &  4 \\
with $\Omega_K$ & 0.059  &  0.048      & 0.044   & 0.058   & 0.147  &  4\\
\hline
Case F     &  0.147       & 0.103   & 0.072     &  0.066     & 0.124  &  2 \\
\hline
\end{tabular}
\caption{$68$\% error in the value of $w$ in the uncorrelated redshift
bins assuming a flat universe. The $z$ range lists the redshift ranges
for the original bins, but decorrelating the covariance matrix results
in a leakage across bins. We show this leakage in the window functions
in Figure~1 for two mock samples  from cases C and E.  The last column
is  the number  of  {\em  independent} EOS  parameters  that could  be
determined to  an accuracy better  than 10\%, consistent with  a prior
study \cite{LinHut}.  For cases B  and C, within brackets, we list the
scatter relative to the mean error of $w_i(z)$ using a moderate number
of random datasets corresponding to the two cases.  For cases C and E,
we also show the errors for  the case where we allow for variations in
curvature with a prior on $\Omega_K$ expected from Planck.}
\label{table1}
\end{center}
\end{table}

Our  results  show  that  it   is  possible  to  determine  {\em  four
independent} EOS parameters  to an accuracy better than  10\%, for the
case of 2300  SNe coupled with high precision  BAO measurements (e.g.,
Case C; Fig.~1).  With cases D and E, two of these four parameters are
determined with  an accuracy  at the level  of 5\%.  Finally,  we note
that the results  of case F show that even with  only 200 SNe combined
with future BAO measurements,  one can constrain three independent EOS
parameters to around 10\%.

The results in Table~1 are for the case of a flat universe, except for
an additional analysis of cases C and E with the inclusion of a Planck
prior on  the curvature~\cite{Smith}. With varying  curvature, we find
that the  errors on $w(z)$  broaden by less  than 3\% relative  to the
errors  with the  flat  universe assumption.  Thus  the assumption  of
flatness, although it very slightly  improves our fits, does not alter
the  general  conclusion  that  one  can  constrain  more  than  three
parameters of the dark energy EOS.

Linder  \& Huterer~\cite{LinHut}  have argued  that future  data (SNe,
CMB, and weak lensing measurements) will lead to a determination of no
more than two  independent parameters of the EOS  to better than 10\%.
They consider  a principal  components analysis of  the EOS  binned at
redshift  intervals  of  0.05,  and  argue that  only  the  first  two
components are  determined to  better than 10\%.  While the  third and
higher  principal  components  are  determined  invidually  with  less
accuracy,  by  combining  multiple components  additional  independent
precise  $w(z_i)$ estimates  can be  achieved.  Our  approach utilizes
much wider  (and uneven) binning  in redshift, thereby allowing  for a
more  robust capture  of dark  energy properties,  and  thus naturally
finding   multiple  independent   parameters.    In  addition,   while
\cite{LinHut}  limited  themselves to  a  Fisher  matrix approach,  we
explicitly generate  mock data sets, accounting  for intrinsic scatter
and  systematic biases  such as  lensing. Even  with the  inclusion of
additional systematic errors, we  find that future data constrain more
than two independent parameters of the dark energy equation of state.

Two-parameter  dark  energy fitting  has  been  incorporated into  the
figure-of-merit (FoM) quantity advocated by the Dark Energy Task Force
\cite{DETF}, which is based on  a two-parameter model for the equation
of state  ($w(z)=w_0+w_a(1-a)$). This provides  a convenient criterion
for  the evaluation  of next-generation  dark energy  surveys.   As an
alternative, more general FoMs, such as the ones discussed in Albrecht
\&  Bernstein~\cite{Albrecht} and  Sullivan  et al.~\cite{sullivan07},
allow for  more than two  parameters.  Since different  FoMs highlight
different aspects of the theory and the data, consideration of a range
of FoMs is warranted.

In summary, we  find that next-generation dark energy  surveys will be
able to constrain three or more independent parameters of the equation
of  state to  an accuracy  better than  10\%. This  is in  contrast to
recent claims in the  literature, and convenventional wisdom, that two
parameters are sufficient in dark energy analyses.  As we enter an era
of precision  measurements, it is  important to avoid  prejudicing our
results with arbitrary  functional forms for the dark  energy. We have
thus proposed a  model-independent, multi-parameter analysis procedure
for fitting  the dark  energy equation of  state, and have  shown that
precision measurements  of the  dark energy can  be expected  with the
next generation of surveys.

\smallskip
We thank  an anonymous referee  and Adam Riess for  valuable comments.
This  work  was  supported  by  LANL IGPP  Astro-1603-07,  NSF  CAREER
AST-0645427  (AC)  and  a  Richard  P. Feynman  Fellowship  from  LANL
(DEH). AC and DEH thank Aspen Center for Physics for hospitality while
this work was completed.



\begin{thebibliography}{99}
\frenchspacing
\bibitem{Rieetal04}
  A.~G.~Riess {\it et al.},
  Astrophys.\ J.\ {\bf 607}, 665 (2004).

\bibitem{Tonryetal03}
  J.~L.~Tonry {\it et al.},
  Astrophys.\ J.\  {\bf 594}, 1 (2003).
  
\bibitem{Eis05}
  D.~J.~Eisenstein {\it et al.},  
  Astrophys.\ J.\  {\bf 633}, 560 (2005).

\bibitem{Spergel06}
  D.~N.~Spergel {\it et al.},
  Astrophys.\ J.\ Suppl.\ {\bf 170}, 377 (2007).


\bibitem{perlmutter-1999}
  A.~Riess {\it et al.}, Astron. J. {\bf 116}, 1009 (1998);
  S.\ Perlmutter {\it et al.},  
  Astrophys.\ J. {\bf 517}, 565 (1999).
    
\bibitem{Wood}
 W.~M.~Wood-Vasey {\it et al.},
  Astrophys.\ J. {\bf 666}, 694 (2007).
    
\bibitem{Rie06}
  A.~G.~Riess {\it et al.},
  Astrophys.\ J.\ {\bf 659}, 98 (2007).

\bibitem{Astier} P. Astier {\it et al.}, 
  Astron. Astrophys. {\bf 447} 31 (2006).
  
\bibitem{Padmanabhan}
  T.~Padmanabhan,
  Phys.\ Rept.\  {\bf 380}, 235 (2003).

\bibitem{CooHut99}
  A.~R.~Cooray and D.~Huterer,
  Astrophys.\ J.\  {\bf 513}, L95 (1999)
  [arXiv:astro-ph/9901097].

\bibitem{ChevPol}
  M.~Chevallier and D.~Polarski, 
  Int. J. Mod. Phys. D {\bf 10}, 213 (2001).

\bibitem{Linder}
	E.V.~Linder
	Phys.\ Rev.\ Lett.\  {\bf 90}, 091301 (2003).
	

\bibitem{Gerke}
  B.~F.~Gerke and G.~Efstathiou,
  Mon.\ Not.\ Roy.\ Astron.\ Soc.\  {\bf 335}, 33 (2002);
  C.~Li, D.~E.~Holz and A.~Cooray,
  Phys.\ Rev.\  D {\bf 75}, 103503 (2007).

\bibitem{HutSta03}
  D.~Huterer and G.~Starkman,
  Phys.\ Rev.\ Lett.\  {\bf 90}, 031301 (2003).

\bibitem{HutCoo}
  D.~Huterer and A.~Cooray,
  Phys.\ Rev.\ D {\bf 71}, 023506 (2005)
  [arXiv:astro-ph/0404062].
  
  \bibitem{Davis07}
  T.~M.~Davis {\it et al.},
  Astrophys.\ J.\  {\bf 666}, 716 (2007)

\bibitem{sullivan07}
  S.~Sullivan, A.~Cooray, and D.~E.~Holz,
  J. Cosmol. Astropart. Phys. 09 (2007) 004

\bibitem{LinHut}
  E.~V.~Linder and D.~Huterer,
  Phys.\ Rev.\  D {\bf 72}, 043509 (2005).

\bibitem{Percival}
  W.~J.~Percival {\it et al.},
  Mon. Not. R. Astron. Soc. {\bf 381}, 1053 (2007)

\bibitem{Seo03}
	H-J.~Seo and D.~J.~Eisenstein,
	Astrophys.\ J.\ {\bf 598}, 720 (2003).
	
\bibitem{Eisen}
  A. Riess (private communication).

\bibitem{kim}
  A.~G.~Kim, E.~V.~Linder, R.~Miquel and N.~Mostek,
  Mon.\ Not.\ Roy.\ Astron.\ Soc.\  {\bf 347}, 909 (2004).

\bibitem{LinHut03}
  E.~V.~Linder and D.~Huterer,
  Phys.\ Rev.\  D {\bf 67}, 081303 (2003).

  \bibitem{Frieman}
  J.~A.~Frieman,
  arXiv:astro-ph/9608068.


  \bibitem{HolzWald98}
  D.~E.~Holz and R.~M.~Wald,
  Phys.\ Rev.\ D {\bf 58}, 063501 (1998).

  \bibitem{Wang00}
  Y.~Wang,
  Astrophys.\ J.\ {\bf 536}, 531 (2000)  

\bibitem{Sarkar}
  D.~Sarkar, A. ~Amblard, D. ~E. ~Holz, and A. ~Cooray,
  Astrophys.\ J.\ {\bf 678}, 1 (2008)
  [arXiv:0710.4143]. 

\bibitem{HolzLinder}
        D.~E.~Holz and E. V. Linder,
        Astrophys.\ J.\ {\bf 631}, 678 (2005).


\bibitem{WangHolzMunshi}
	Y.~Wang, D.~E.~Holz, and D.~Munshi,
	Astrophys.\ J.\ {\bf 572}, L15 (2002).

\bibitem{Tegmark04}  
	M.~Tegmark et al.,
	Astrophys.\ J.\ {\bf 606}, 702 (2004).


\bibitem{key01}
W. L. ~Freedman et al. ,
Astrophys. \ J. \ {\bf 553}, 47 (2001).


\bibitem{Wang}
  Y.~Wang and P.~Mukherjee,
  Phys. Rev. D {\bf 76}, 103533 (2007).

\bibitem{Smith}
K.~M.~Smith, W.~Hu, and M.~Kaplinghat, Phys.\ Rev.\ D {\bf 74}, 123002 (2006).

\bibitem{DETF}
  A.~Albrecht {\it et al.},
  arXiv:astro-ph/0609591 (2006).

\bibitem{Albrecht}
  A.~Albrecht and G.~Bernstein,
  Phys.\ Rev.\  D {\bf 75}, 103003 (2007).
  

\end{thebibliography}
\end{document}